\documentstyle[12pt,epsf,epsfig]{article}   
\renewcommand{\baselinestretch}{1.5}

 1

\catcode`\@=11 
\makeatletter
\def\@seccntformat#1{\csname the#1\endcsname.\hskip 1em}

\makeatother

\catcode`@=11
\def\chkspace{%
  \relax   
  \begingroup\ifhmode\aftergroup\dochksp@ce\fi\endgroup}
\def\dochksp@ce{%
  \unskip              
  \futurelet\chkspct@k\d@chkspc  
}
\def\d@chkspc{%
  \let\nxtsp@ce=\relax
  \ifx\chkspct@k.\else     
    \ifx\chkspct@k,\else
      \ifx\chkspct@k;\else
        \ifx\chkspct@k!\else
          \ifx\chkspct@k?\else
            \ifx\chkspct@k:\else
              \ifx\chkspct@k)\else
              \ifx\chkspct@k(\else
                \ifx\chkspct@k]\else
                  \ifx\chkspct@k-\else
                    \ifx\chkspct@k\egroup\else  
                      \let\nxtsp@ce=\put@space  
                    \fi
                  \fi
                \fi
              \fi
              \fi
            \fi
          \fi
        \fi
      \fi
    \fi
  \fi
  \nxtsp@ce
}
\def\put@space{$\;$}
\catcode`@=12

\def\ra{{$\rightarrow$}\chkspace}
\def\etal{{\it et al.}\chkspace}

\def\eg{{\it eg.}\chkspace}

\def\ep{{e$^+$e$^-$}\chkspace}

\def\gluino{\relax\ifmmode \tilde{g} \else $\tilde{g}$ \fi\chkspace}

\def\qq{\relax\ifmmode q\overline{q}
\else $q\overline{q}$ \fi\chkspace}

\def\bb{\relax\ifmmode b\bar{b}
       \else $b\bar{b}$ \fi\chkspace}
\def\ccrm{\relax\ifmmode {\rm c}\bar{\rm c}
       \else ${\rm c}\bar{\rm c}$ \fi\chkspace}

\def\tt{\relax\ifmmode {\rm t}\bar{\rm t}
       \else ${\rm t}\bar{\rm t}$ \fi\chkspace}
\def\ss{\relax\ifmmode {\rm s}\bar{\rm s}
       \else ${\rm s}\bar{\rm s}$ \fi\chkspace}
\def\uu{\relax\ifmmode {\rm u}\bar{\rm u}
       \else ${\rm u}\bar{\rm u}$ \fi\chkspace}
\def\dd{\relax\ifmmode {\rm d}\bar{\rm d}
       \else ${\rm d}\bar{\rm d}$ \fi\chkspace}

\def\qqg{\relax\ifmmode q\overline{q}g
\else $q\overline{q}g$ \fi\chkspace}
\def\bbg{\relax\ifmmode b\overline{b}g
\else $b\overline{b}g$ \fi\chkspace}

\def\afb{\relax\ifmmode A_{FB} \else
{{$A_{FB}$}}\fi\chkspace}
\def\afbb{\relax\ifmmode A_{FB}^b \else
{{$A_{FB}^b$}}\fi\chkspace}
\def\pafb{\relax\ifmmode \tilde{A}_{FB} \else
{{$\tilde{A}_{FB}$}}\fi\chkspace}
\def\pafbb{\relax\ifmmode \tilde{A}_{FB}^b \else
{{$\tilde{A}_{FB}^b$}}\fi\chkspace}

\def\pafbzo{\relax\ifmmode \tilde{A}_{FB}|_{O(0)} \else
{{$\tilde{A}_{FB}|_{O(0)}$}}\fi\chkspace}
\def\pafbfo{\relax\ifmmode \tilde{A}_{FB}|_{\oalp} \else
{{$\tilde{A}_{FB}|_{\oalp}$}}\fi\chkspace}
\def\pafbso{\relax\ifmmode \tilde{A}_{FB}|_{\oalpsq} \else
{{$\tilde{A}_{FB}|_{\oalpsq}$}}\fi\chkspace}
\def\pafbto{\relax\ifmmode \tilde{A}_{FB}|_{\oalpc} \else
{{$\tilde{A}_{FB}|_{\oalpc}$}}\fi\chkspace}

\def\pafbbzo{\relax\ifmmode \tilde{A}_{FB}^b|_{O(0)} \else
{{$\tilde{A}_{FB}^b|_{O(0)}$}}\fi\chkspace}
\def\pafbbfo{\relax\ifmmode \tilde{A}_{FB}^b|_{\oalp} \else
{{$\tilde{A}_{FB}^b|_{\oalp}$}}\fi\chkspace}
\def\pafbbso{\relax\ifmmode \tilde{A}_{FB}^b|_{\oalpsq} \else
{{$\tilde{A}_{FB}^b|_{\oalpsq}$}}\fi\chkspace}
\def\pafbbto{\relax\ifmmode \tilde{A}_{FB}^b|_{\oalpc} \else
{{$\tilde{A}_{FB}^b|_{\oalpc}$}}\fi\chkspace}

\def\afbo0{\tilde{A}_{FB}|_{O(0)}}
\def\afbo1{\tilde{A}_{FB}|_{\oalp}}
\def\afbo2{\tilde{A}_{FB}|_{\oalpsq}}
\def\afbo3{\tilde{A}_{FB}|_{\oalpc}}

\def\lam{\relax\ifmmode \Lambda_{\overline{MS}}
       \else {{$\Lambda_{\overline{MS}}$}}\fi\chkspace}
\def\lamuds{\relax\ifmmode \Lambda^{(3)}_{\overline{MS}}
       \else {{$\Lambda^{(3)}_{\overline{MS}}$}}\fi\chkspace}
\def\lamudsc{\relax\ifmmode \Lambda^{(4)}_{\overline{MS}}
       \else $\Lambda^{(4)}_{\overline{MS}}$\fi\chkspace}
\def\lamudscb{\relax\ifmmode \Lambda^{(5)}_{\overline{MS}}
       \else $\Lambda^{(5)}_{\overline{MS}}$\fi\chkspace}

\def\alp{\relax\ifmmode \alpha_s\else $\alpha_s$\fi\chkspace}
\def\alpbar{\relax\ifmmode \bar{\alpha_s}
       \else $\bar{\alpha_s}$\fi\chkspace}
\def\alpmz{\relax\ifmmode \alpha_s(M_Z)\else $\alpha_s(M_Z)$\fi\chkspace}
\def\alpmzsq{\relax\ifmmode \alpha_s(M_Z^2)
       \else $\alpha_s(M_Z^2)$\fi\chkspace}

\def\oalp{\relax\ifmmode O(\alpha_s)\else{{O($\alpha_s$)}}\fi\chkspace}
\def\oalpsq{\relax\ifmmode O(\alpha_s^2)
           \else{{O($\alpha_s^2$)}}\fi\chkspace}
\def\oalpc{\relax\ifmmode O(\alpha_s^3)
           \else{{O($\alpha_s^3$)}}\fi\chkspace}
\def\oalpf{\relax\ifmmode O(\alpha_s^4)
           \else{{O($\alpha_s^4$)}}\fi\chkspace}

\def\rb{\relax\ifmmode R_3^b/R_3^{all}
           \else{{$R_3^b/R_3^{all}$}}\fi\chkspace}
\def\rc{\relax\ifmmode R_3^c/R_3^{all}
           \else{{$R_3^c/R_3^{all}$}}\fi\chkspace}
\def\ruds{\relax\ifmmode R_3^{uds}/R_3^{all}
           \else{{$R_3^{uds}/R_3^{all}$}}\fi\chkspace}
\def\ri{\relax\ifmmode R_3^i/R_3^{all}
           \else{{$R_3^i/R_3^{all}$}}\fi\chkspace}
\def\rj{\relax\ifmmode R_3^j/R_3^{all}
           \else{{$R_3^j/R_3^{all}$}}\fi\chkspace}
\def\alpi{\relax\ifmmode \alpha^i_s/\alpha^{all}_s
           \else{{$\alpha^i_s/\alpha^{all}_s$}}\fi\chkspace}

\def\mbz{\relax\ifmmode m_b(M_Z)
           \else{{$m_b(M_Z)$}}\fi\chkspace}
\def\mbb{\relax\ifmmode m_b(M_b)
           \else{{$m_b(M_b)$}}\fi\chkspace}

\def\plb{Phys. Lett.\chkspace}

\def\prl{Phys. Rev. Lett.\chkspace}
\def\prd{Phys. Rev.\chkspace}

\def\z0{{$Z^0$}\chkspace}
\def\Dst{\relax\ifmmode {\rm D}^* \else {D$^*$}\fi\chkspace}
\def\Dpl{\relax\ifmmode {\rm D}^+ \else {D$^+$}\fi\chkspace}
\def\D0{\relax\ifmmode {\rm D}^0 \else {D$^0$}\fi\chkspace}
\def\Kst{\relax\ifmmode {\rm K}^* \else {K$^*$}\fi\chkspace}
\def\K0{\relax\ifmmode {\rm K}^0_s \else {K$^0_s$}\fi\chkspace}
\def\Kpl{\relax\ifmmode {\rm K}^+ \else {K$^+$}\fi\chkspace}
\def\Kstz{\relax\ifmmode {\rm K}^{*0} \else {K$^{*0}$}\fi\chkspace}

\def\beq{\begin{equation}}
\def\eeq{\end{equation}}
\def\bea{\begin{eqnarray}}
\def\eea{\end{eqnarray}}

\begin{document}

\thispagestyle{empty}
\begin{flushright}
{\renewcommand{\baselinestretch}{.75}
  SLAC--PUB--8155\\
June 1999\\
}
\end{flushright}

\vskip 1truecm
 
\begin{center}
{\large\bf
 STUDY OF THE STRUCTURE\\
 OF e$^+$e$^-$ \ra \bbg  EVENTS AND IMPROVED LIMITS\\
 ON THE ANOMALOUS CHROMOMAGNETIC\\ 
COUPLING OF THE $b$-QUARK$^*$
}

\end{center}
 
 
\begin{center}
 {\bf The SLD Collaboration$^{**}$}\\
Stanford Linear Accelerator Center \\
Stanford University, Stanford, CA~94309
\end{center}
 
\vspace{1cm}
 
\begin{center}
{\bf ABSTRACT }
\end{center}
 
\noindent
The structure of \ep \ra $\bbg$ events was studied using $Z^0$ 
decays recorded in the SLD experiment at SLAC. Three-jet final states were 
selected 
and the CCD-based vertex detector was used to identify two of the jets as  
$b$ or $\overline{b}$. Distributions of the gluon energy and polar angle 
were measured over the full kinematic range, and compared with
perturbative QCD predictions. The energy distribution is
 potentially sensitive to 
an anomalous $b$ chromomagnetic moment $\kappa$. We measured $\kappa$ 
to be consistent with zero and set limits on its value, 
$-0.11<\kappa<0.08$ at 95\%~c.l. (preliminary). 

\vskip 1truecm

\vfill
\noindent
Contributed to: the International Europhysics Conference on High Energy Physics,
15-21 July 1999, Tampere, Finland; Ref. 1\_182, and to the XIXth International 
Symposium on Lepton and Photon Interactions, August 9-14 1999, Stanford, USA.

{\footnotesize
$^*$ Work supported by Department of Energy contract DE-AC03-76SF00515 (SLAC).}

\eject

\noindent
The observation of $e^+e^-$ annihilation into final states containing three 
hadronic jets, and their interpretation in terms of the 
process $e^+e^-\rightarrow\qqg$~\cite{threejets}, provided the first direct 
evidence for the existence of the gluon, the gauge boson of the theory of 
strong interactions, Quantum Chromodynamics (QCD). 
In subsequent studies the jets 
were usually energy ordered, and the lowest-energy jet was assigned as the gluon; 
this is correct roughly 80\% of the time, but preferentially
selects low-energy gluons. If the gluon jet could be tagged explicitly, event-by-event, the
full kinematic range of gluon energies could be explored, and
more detailed tests of QCD could be performed~\cite{BO}. Due to advances in 
vertex-detection this is now possible using \ep \ra 
$\bbg$ events. The large mass 
and relatively long lifetime, $\sim$ 1.5 ps, of the leading $B$ hadron in $b$-quark 
jets~\cite{quark} lead to decay signatures which 
distinguish them from lighter-quark ($u$, $d$, $s$ or $c$) and gluon jets.
We used our original (1992-5) and upgraded (1996-8)
CCD vertex detectors~\cite{VXD2,VXD3} 
to identify in each event the two jets that contain the $B$ 
hadrons, and hence to tag the gluon jet. This allowed us to measure the gluon
energy and polar-angle distributions over the full kinematic range.

Additional motivation to study the \bbg system has been provided by 
measurements involving inclusive \z0\ra\bb decays.
Several reported determinations~\cite{electrow} of   
$R_b$ = $\Gamma(Z^0\rightarrow$\bb)/$\Gamma(Z^0\rightarrow$\qq) 
and the \z0-$b$ parity-violating coupling parameter, $A_b$, 
differed from Standard Model (SM) expectations at the few standard deviation level. 
Since one expects new high-mass-scale 
dynamics to couple to the massive third-generation fermions, these 
measurements aroused considerable interest and speculation. 
We have therefore investigated in detail the
strong-interaction dynamics of the $b$-quark.
We have compared the strong coupling of the gluon to
$b$-quarks with that to light- and charm-quarks~\cite{sldflav}, as well as
tested parity (P) and charge$\oplus$parity (CP) conservation at the \bbg 
vertex~\cite{sldsymm}. We have also studied   
the structure of \bbg events via the 
distributions of the gluon 
energy and polar angle with respect to (w.r.t.) the beamline~\cite{sldbbg};
here we present a preliminary update of these measurements using a data
sample more than 3 times larger than in our earlier study.
We compare these results with perturbative QCD predictions, including a
recent calculation at next-to-leading order (NLO) which takes quark mass
effects into account~\cite{arnd}. 

In QCD the chromomagnetic moment of the $b$ quark is induced at the 
one-loop level and is of order $\alpha_s$/$\pi$. 
A more general $\bbg$ Lagrangian term with a modified coupling~\cite{tom1}
may be written:  
\begin{equation}
{\cal L}^{b\overline{b}g} =  g_s\overline{b}T_a \{ \gamma_{\mu} + 
\frac{i\sigma_{\mu\nu}k^{\nu}}{2m_b}(\kappa - i \tilde{\kappa}\gamma_5)\} 
bG_a^{\mu},\label{lag}
\end{equation}
\noindent
where 
$\kappa$ and $\tilde{\kappa}$ parameterize the anomalous chromomagnetic and chromoelectric
moments, respectively, which might arise from physics beyond the SM. 
The effects of the chromoelectric moment are sub-leading w.r.t. those of the
chromomagentic moment, so for convenience we set 
$\tilde{\kappa}$ to zero. A non-zero $\kappa$
would modify~\cite{tom1} the gluon energy distribution in $\bbg$ events relative to the
standard QCD case. Hence we have used our larger data sample to set improved limits on 
$\kappa$.

We used hadronic
decays of $Z^0$ bosons produced by $e^+e^-$ annihilations at the SLAC Linear
Collider (SLC) which were recorded in the SLC Large Detector (SLD)~\cite{SLD}. 
The criteria for selecting \z0 decays, and the charged tracks used 
for flavor-tagging, are described in~\cite{sldflav,dervan}.
We applied the JADE algorithm~\cite{jade} to define jets, using a 
scaled-invariant-mass criterion  
$y_{cut}$ = 0.02. Events classified as 3-jet states were retained if all three
jets were well contained within the barrel tracking system, with polar angle $|\cos
\theta_{jet}|$ $\leq$ 0.71. From our 1993-98 data samples, comprising
roughly 500,000 hadronic \z0 decays, 126,871 events were selected. 
In order to improve the energy resolution 
the jet energies were rescaled kinematically according to the angles
between the jet axes, assuming energy and momentum conservation and massless
kinematics~\cite{sldsymm}. 
The jets were then labelled in order of energy such that $E_1 > E_2 > E_3$.

Charged tracks with a large transverse impact parameter w.r.t. the
measured interaction point (IP) were used to tag \bbg events~\cite{sldflav}. The
resolution on the impact parameter, projected in the plane
normal to
the beamline, $d$, is $\sigma_d$ = 
11$\oplus$70/($p_{\perp}\sqrt{\sin\theta}$) (1993-5)
and 8$\oplus$33/($p_{\perp}\sqrt{\sin\theta}$) (1996-8)
$\mu$m, where $p_{\perp}$ is the track transverse momentum in GeV/c, and
$\theta$ the polar angle, w.r.t. the beamline. 
The flavour tag was based on the number of tracks per jet, $N_{sig}^{jet}$, 
with $d/\sigma_d\geq3$~\cite{sldbbg}. 
Events were retained in which exactly two jets were $b$-tagged 
by requiring each to have $N_{sig}^{jet}\geq2$, and in which
the remaining jet had $N_{sig}^{jet}<2$ and was hence tagged as the gluon; 
8196 events were selected. The 
efficiency for selecting true \bbg events is 12$\%$.
This was estimated using a simulated event sample generated with  
JETSET 7.4 \cite{jetset}, with parameter values
tuned to hadronic $e^+e^-$ annihilation data \cite{tuning}, combined with a
simulation of
$B$-decays tuned to $\Upsilon$(4S) data \cite{bdecay} and a simulation of the
detector.
The efficiency peaks at about 15\% for 15 GeV gluons. Lower-energy
gluon jets are sometimes merged 
with the parent $b$-jet by the jet-finder. At higher gluon energies the 
the correspondingly lower-energy $b$-jets are harder to tag, and 
there is also a higher probability of losing a jet outside the detector acceptance. 

For the selected event sample, Fig.~1 shows the  $N_{sig}^{jet}$
distributions separately for jets 1, 2 and 3.
In about 16\% of cases the gluon-tagged jet is not the lowest-energy jet (jet 3).
The simulated contributions from
true gluons are indicated~\cite{bbbb}
and the estimated gluon purities~\cite{bbbb} are listed in Table~\ref{dtagpure}.
The inclusive gluon purity of the tagged-jet sample is 93\%. 
With this sample we formed the distributions of two gluon-jet observables, 
the scaled energy $x_g = 2E_{\rm{gluon}}/\sqrt{s}$, and the polar angle 
w.r.t. the beamline, $\theta_g$. The distributions 
are shown in Fig.~2. The simulation is also shown; it reproduces the data. 

The backgrounds were estimated using the simulation and
are of three types: non-$\bb$ events, 
$\bb$ but non-$\bbg$ events, and true $\bbg$ events
in which the gluon jet was mis-tagged as a $b$-jet. These 
are shown in Fig.~2. The non-$\bb$ events ($\sim$ 5$\%$ of the \bbg sample) are 
mainly $c\overline{c}g$ events, 
92\% of which had the gluon correctly tagged. There is a small
contribution ($\sim$ 0.1$\%$ of the \bbg sample) from light-quark events. 
The dominant background is formed by \bb but non-$\bbg$ events. These are true 
$\bb$ events which were not
classified as 3-jet events at the parton level,
but which were mis-reconstructed and tagged as 3-jet $\bbg$ events 
in the detector using the same jet algorithm and $y_{cut}$ value.
This arises from the broadening of the particle flow around the
original $b$ and $\overline{b}$ directions due to hadronisation and the high-transverse-momentum 
$B$-decay products, causing the jet-finder to reconstruct a `fake' third 
jet, which is almost
always assigned as the gluon. The population of such fake gluon jets peaks at low energy
(Fig.~2(a)), as expected.
Mis-tagged events comprise less than 1\% of the \bbg sample.

The distributions were corrected to obtain the true gluon
distributions $D^{true}(X)$ by applying a bin-by-bin procedure:
$D^{true}(X) = C(X)\;(D^{raw}(X)-B(X))$,
where $X$ = $x_g$ or cos$\theta_g$, $D^{raw}(X)$ 
is the raw distribution,
$B(X)$ is the background contribution, and
$C(X) \equiv D^{true}_{MC}(X)/D^{recon}_{MC}(X)$
is a correction that accounts for the
efficiency for accepting true \bbg events into the tagged sample, as well as
for bin-to-bin migrations caused by hadronisation, 
the resolution of the detector, and bias of the jet-tagging technique.
Here $D^{true}_{MC}(X)$ is the true distribution for MC-generated \bbg events,
and $D^{recon}_{MC}(X)$ is the resulting distribution after full
simulation of the detector and application of the same analysis procedure
as applied to the data.

As a cross-check, an alternative correction procedure was employed 
in which bin-to-bin migrations,
which can be as large as 20\%, were explicitly taken into account:
$D^{true}(X_i) = M(X_i,X_j)\;(D^{raw}(X_j)-B(X_j))/\epsilon(X_i)$,
with the unfolding matrix $M(X_i,X_j)$ defined by
$D^{true}_{MC}(X_i) = M(X_i,X_j) D^{recon}_{MC}(X_j)$,
where true $\bbg$ events generated in bin $i$ may, 
after reconstruction, be accepted into the tagged sample in bin $j$.
$\epsilon(X)$ is the
efficiency for accepting \bbg events in bin $i$ into the tagged sample.
The resulting distributions of $x_g$ and cos$\theta_g$ 
are statistically indistinguishable from the respective 
distributions yielded by the bin-by-bin method.

The fully-corrected distributions are shown in Fig.~3. Since, in an
earlier study~\cite{sldflav}, we verified that the overall rate of \bbg-event production is
consistent with QCD expectations, we normalised the gluon distributions to
unit area and we study further the distribution shapes. 
The $x_g$ distribution rises, peaks around $x_g$ $\sim$ 0.15,
and decreases towards zero as $x_g$ $\rightarrow$ 1. 
The peak is a kinematic artifact of the jet algorithm, 
which ensures that gluon jets are reconstructed with a non-zero energy
which depends on the $y_c$ value. The cos$\theta_g$ distribution is flat.

We have considered sources of systematic uncertainty that potentially affect
our results. These may be divided into uncertainties in
modelling the detector and uncertainties in the underlying physics modelling. 
To estimate the first case we systematically varied the track and
event selection requirements, as well as the tracking efficiency~\cite{sldflav,dervan}. 
In the second case parameters used in our simulation,
relating mainly to the production and decay of
charm and bottom hadrons,  
were varied within their measurement errors~\cite{dervan}. 
For each variation the data were recorrected to
derive new $x_g$ and cos$\theta_g$ distributions, and the deviation w.r.t. the
standard case was assigned as a systematic uncertainty. 
None of the variations affects our conclusions.
All uncertainties were conservatively assumed to be
uncorrelated and were added in 
quadrature in each bin of $x_g$ and cos$\theta_g$.

We compared the data with perturbative QCD predictions for
the same jet algorithm and $y_c$ value. We used leading-order (LO) and
NLO results based on recent calculations~\cite{arnd} 
in which quark mass effects were explicitly taken into
account; a $b$-mass value of $m_b(m_Z)=3$GeV/$c^2$ was used~\cite{bmass}.
We also derived these distributions using the `parton 
shower' (PS) implemented in JETSET. This is equivalent to a
calculation in which all leading, and a subset of next-to-leading, 
ln$y_c$ terms are resummed to all orders in \alp. In physical terms this
allows events to be generated with multiple orders of parton radiation,
in contrast to the maximum number of 3 (4) partons allowed in the LO
~(NLO) calculations, respectively. Configurations with $\geq3$ partons are
relevant to the observables considered here since they may be resolved 
as 3-jet events by the jet-finding algorithm.

These predictions are shown in Fig.~3. 
The calculations reproduce the measured cos$\theta_g$ distribution,
which is clearly insensitive to the details of higher-order soft parton
emission. For $x_g$, although the LO 
calculation reproduces the main features of the shape of the distribution,
it yields too few events in the region $0.2<x_g<0.5$, and too many events
for $x_g<0.1$ and $x_g>0.5$. The NLO calculation is noticeably better, but
also shows a deficit for $0.2<x_g<0.4$. The PS calculation describes the
data across the full $x_g$ range.
These results suggest that multiple
orders of parton radiation need to be included, in
agreement with our earlier measurements of jet energy distributions using 
flavor-inclusive \z0 decays~\cite{hwang}. 
We also investigated LO and NLO predictions based on  
matrix elements implemented in JETSET which assume massless quarks.
The resulting distributions are practically indistinguishable from
the massive ones, even though the large $b$-mass has been seen~\cite{bmass}
to affect the \bbg event rate at the level of  
5\%. The effect of varying $\alpha_s$ within the world-average
range is similarly small.

We conclude that perturbative QCD in the PS approximation accurately reproduces 
the gluon distributions in \bbg events. However, it is interesting to
consider the extent to which anomalous chromomagnetic contributions are
allowed.
The Lagrangian represented by Eq.~\ref{lag} yields a model that is
non-renormalisable.  Nevertheless tree-level predictions can be
derived \cite{tom1} and used for a `straw man' comparison with QCD. 
For illustration, the effect of a large anomalous moment, 
$\kappa=0.75$, on the shape of the $x_g$ distribution is shown in 
Fig.~3(a); there is a clear depletion of events in the region $x_g < 0.5$
and a corresponding enhancement for $x_g \geq 0.5$. By contrast the 
shape of the cos$\theta_g$ distribution is relatively unchanged (not shown), 
even by such a large $\kappa$ value.
In each bin of the
$x_g$ distribution, we parametrised the leading-order effect of an anomalous
chromomagnetic moment
and added it to the PS calculation to arrive at an effective QCD
prediction including the anomalous moment at leading-order.
A $\chi^2$ minimisation fit was performed to the
data with $\kappa$ as a free parameter,
yielding 
$\kappa = -0.011 \pm0.048{\rm (stat.)}^{+0.013}_{-0.003}{\rm (syst.)}$,
which is consistent with zero within the errors,
with a $\chi^2$ of 17.8 for 9 degrees of freedom. 
The distribution corresponding
to this fit is indistinguishable from the PS prediction 
(Fig.~3(a)) and is not shown.
Our result corresponds to 95$\%$  confidence-level (c.l.) upper limits of 
$-0.11 < \kappa < 0.08$ (preliminary). 
 
In conclusion, we used the precise SLD tracking system to tag the gluon in
3-jet $e^+e^-\rightarrow Z^0\rightarrow\bbg$ events. We studied the structure
of these events in terms of the scaled gluon energy and polar angle, measured 
across the full kinematic range.
We compared our data with perturbative QCD predictions, and found that
the effect of the $b$-mass on the shapes of the distributions is small, that
beyond-LO QCD contributions are needed to describe the energy
distribution, and that 
the parton shower prediction agrees best with the data. We also
investigated an anomalous $b$-quark chromomagnetic moment, $\kappa$, which 
would affect the shape of the energy distribution. We set preliminary 
95$\%$ c.l. limits of 
$-0.11 < \kappa < 0.08$. 

\vskip .5truecm

We thank the personnel of the SLAC accelerator department and the
technical
staffs of our collaborating institutions for their outstanding efforts
on our behalf. We thank A.~Brandenburg, P.~Uwer and T.~Rizzo for many helpful
discussions and for their calculational efforts on our behalf.

\vskip 1truecm

\vbox{
\footnotesize\renewcommand{\baselinestretch}{1}
\noindent
 This work was supported by Department of Energy
  contracts:
  DE-FG02-91ER40676 (BU),
  DE-FG03-91ER40618 (UCSB),
  DE-FG03-92ER40689 (UCSC),
  DE-FG03-93ER40788 (CSU),
  DE-FG02-91ER40672 (Colorado),
  DE-FG02-91ER40677 (Illinois),
  DE-AC03-76SF00098 (LBL),
  DE-FG02-92ER40715 (Massachusetts),
  DE-FC02-94ER40818 (MIT),
  DE-FG03-96ER40969 (Oregon),
  DE-AC03-76SF00515 (SLAC),
  DE-FG05-91ER40627 (Tennessee),
  DE-FG02-95ER40896 (Wisconsin),
  DE-FG02-92ER40704 (Yale);
  National Science Foundation grants:
  PHY-91-13428 (UCSC),
  PHY-89-21320 (Columbia),
  PHY-92-04239 (Cincinnati),
  PHY-95-10439 (Rutgers),
  PHY-88-19316 (Vanderbilt),
  PHY-92-03212 (Washington);
  the UK Particle Physics and Astronomy Research Council
  (Brunel, Oxford and RAL);
  the Istituto Nazionale di Fisica Nucleare of Italy
  (Bologna, Ferrara, Frascati, Pisa, Padova, Perugia);
  the Japan-US Cooperative Research Project on High Energy Physics
  (Nagoya, Tohoku);
  and the Korea Science and Engineering Foundation (Soongsil).}
  

\vfill
\eject

\section*{$^{**}$List of Authors}
%
%
%
\begin{center}
\def\iADEL{$^{(1)}$}
\def\iAOMORI{$^{(2)}$}
\def\iBOLO{$^{(3)}$}
\def\iBRI{$^{(4)}$}
\def\iBRUN{$^{(5)}$}
\def\iBU{$^{(6)}$}
\def\iCINC{$^{(7)}$}
\def\iCOLO{$^{(8)}$}
\def\iCOLU{$^{(9)}$}
\def\iCSU{$^{(10)}$}
\def\iFERR{$^{(11)}$}
\def\iFRAS{$^{(12)}$}
\def\iILLI{$^{(13)}$}
\def\iJHU{$^{(14)}$}
\def\iLBL{$^{(15)}$}
\def\iLTU{$^{(16)}$}
\def\iMASS{$^{(17)}$}
\def\iMISSI{$^{(18)}$}
\def\iMIT{$^{(19)}$}
\def\iMOSCOW{$^{(20)}$}
\def\iNAGO{$^{(21)}$}
\def\iOREG{$^{(22)}$}
\def\iOXF{$^{(23)}$}
\def\iPADO{$^{(24)}$}
\def\iPERU{$^{(25)}$}
\def\iPISA{$^{(26)}$}
\def\iRAL{$^{(27)}$}
\def\iRUTG{$^{(28)}$}
\def\iSLAC{$^{(29)}$}
\def\iSOGA{$^{(30)}$}
\def\iSOONG{$^{(31)}$}
\def\iTENN{$^{(32)}$}
\def\iTOHO{$^{(33)}$}
\def\iUCSB{$^{(34)}$}
\def\iUCSC{$^{(35)}$}
\def\iUVIC{$^{(36)}$}
\def\iVAND{$^{(37)}$}
\def\iWASH{$^{(38)}$}
\def\iWISC{$^{(39)}$}
\def\iYALE{$^{(40)}$}

  \baselineskip=.75\baselineskip  
\mbox{Kenji  Abe\unskip,\iNAGO}
\mbox{Koya Abe\unskip,\iTOHO}
\mbox{T. Abe\unskip,\iSLAC}
\mbox{I.Adam\unskip,\iSLAC}
\mbox{T.  Akagi\unskip,\iSLAC}
\mbox{N. J. Allen\unskip,\iBRUN}
\mbox{W.W. Ash\unskip,\iSLAC}
\mbox{D. Aston\unskip,\iSLAC}
\mbox{K.G. Baird\unskip,\iMASS}
\mbox{C. Baltay\unskip,\iYALE}
\mbox{H.R. Band\unskip,\iWISC}
\mbox{M.B. Barakat\unskip,\iLTU}
\mbox{O. Bardon\unskip,\iMIT}
\mbox{T.L. Barklow\unskip,\iSLAC}
\mbox{G. L. Bashindzhagyan\unskip,\iMOSCOW}
\mbox{J.M. Bauer\unskip,\iMISSI}
\mbox{G. Bellodi\unskip,\iOXF}
\mbox{R. Ben-David\unskip,\iYALE}
\mbox{A.C. Benvenuti\unskip,\iBOLO}
\mbox{G.M. Bilei\unskip,\iPERU}
\mbox{D. Bisello\unskip,\iPADO}
\mbox{G. Blaylock\unskip,\iMASS}
\mbox{J.R. Bogart\unskip,\iSLAC}
\mbox{G.R. Bower\unskip,\iSLAC}
\mbox{J. E. Brau\unskip,\iOREG}
\mbox{M. Breidenbach\unskip,\iSLAC}
\mbox{W.M. Bugg\unskip,\iTENN}
\mbox{D. Burke\unskip,\iSLAC}
\mbox{T.H. Burnett\unskip,\iWASH}
\mbox{P.N. Burrows\unskip,\iOXF}
\mbox{A. Calcaterra\unskip,\iFRAS}
\mbox{D. Calloway\unskip,\iSLAC}
\mbox{B. Camanzi\unskip,\iFERR}
\mbox{M. Carpinelli\unskip,\iPISA}
\mbox{R. Cassell\unskip,\iSLAC}
\mbox{R. Castaldi\unskip,\iPISA}
\mbox{A. Castro\unskip,\iPADO}
\mbox{M. Cavalli-Sforza\unskip,\iUCSC}
\mbox{A. Chou\unskip,\iSLAC}
\mbox{E. Church\unskip,\iWASH}
\mbox{H.O. Cohn\unskip,\iTENN}
\mbox{J.A. Coller\unskip,\iBU}
\mbox{M.R. Convery\unskip,\iSLAC}
\mbox{V. Cook\unskip,\iWASH}
\mbox{R. Cotton\unskip,\iBRUN}
\mbox{R.F. Cowan\unskip,\iMIT}
\mbox{D.G. Coyne\unskip,\iUCSC}
\mbox{G. Crawford\unskip,\iSLAC}
\mbox{C.J.S. Damerell\unskip,\iRAL}
\mbox{M. N. Danielson\unskip,\iCOLO}
\mbox{M. Daoudi\unskip,\iSLAC}
\mbox{N. de Groot\unskip,\iBRI}
\mbox{R. Dell'Orso\unskip,\iPERU}
\mbox{P.J. Dervan\unskip,\iBRUN}
\mbox{R. de Sangro\unskip,\iFRAS}
\mbox{M. Dima\unskip,\iCSU}
\mbox{A. D'Oliveira\unskip,\iCINC}
\mbox{D.N. Dong\unskip,\iMIT}
\mbox{M. Doser\unskip,\iSLAC}
\mbox{R. Dubois\unskip,\iSLAC}
\mbox{B.I. Eisenstein\unskip,\iILLI}
\mbox{V. Eschenburg\unskip,\iMISSI}
\mbox{E. Etzion\unskip,\iWISC}
\mbox{S. Fahey\unskip,\iCOLO}
\mbox{D. Falciai\unskip,\iFRAS}
\mbox{C. Fan\unskip,\iCOLO}
\mbox{J.P. Fernandez\unskip,\iUCSC}
\mbox{M.J. Fero\unskip,\iMIT}
\mbox{K.Flood\unskip,\iMASS}
\mbox{R. Frey\unskip,\iOREG}
\mbox{J. Gifford\unskip,\iUVIC}
\mbox{T. Gillman\unskip,\iRAL}
\mbox{G. Gladding\unskip,\iILLI}
\mbox{S. Gonzalez\unskip,\iMIT}
\mbox{E. R. Goodman\unskip,\iCOLO}
\mbox{E.L. Hart\unskip,\iTENN}
\mbox{J.L. Harton\unskip,\iCSU}
\mbox{A. Hasan\unskip,\iBRUN}
\mbox{K. Hasuko\unskip,\iTOHO}
\mbox{S. J. Hedges\unskip,\iBU}
\mbox{S.S. Hertzbach\unskip,\iMASS}
\mbox{M.D. Hildreth\unskip,\iSLAC}
\mbox{J. Huber\unskip,\iOREG}
\mbox{M.E. Huffer\unskip,\iSLAC}
\mbox{E.W. Hughes\unskip,\iSLAC}
\mbox{X.Huynh\unskip,\iSLAC}
\mbox{H. Hwang\unskip,\iOREG}
\mbox{M. Iwasaki\unskip,\iOREG}
\mbox{D. J. Jackson\unskip,\iRAL}
\mbox{P. Jacques\unskip,\iRUTG}
\mbox{J.A. Jaros\unskip,\iSLAC}
\mbox{Z.Y. Jiang\unskip,\iSLAC}
\mbox{A.S. Johnson\unskip,\iSLAC}
\mbox{J.R. Johnson\unskip,\iWISC}
\mbox{R.A. Johnson\unskip,\iCINC}
\mbox{T. Junk\unskip,\iSLAC}
\mbox{R. Kajikawa\unskip,\iNAGO}
\mbox{M. Kalelkar\unskip,\iRUTG}
\mbox{Y. Kamyshkov\unskip,\iTENN}
\mbox{H.J. Kang\unskip,\iRUTG}
\mbox{I. Karliner\unskip,\iILLI}
\mbox{H. Kawahara\unskip,\iSLAC}
\mbox{Y. D. Kim\unskip,\iSOGA}
\mbox{M.E. King\unskip,\iSLAC}
\mbox{R. King\unskip,\iSLAC}
\mbox{R.R. Kofler\unskip,\iMASS}
\mbox{N.M. Krishna\unskip,\iCOLO}
\mbox{R.S. Kroeger\unskip,\iMISSI}
\mbox{M. Langston\unskip,\iOREG}
\mbox{A. Lath\unskip,\iMIT}
\mbox{D.W.G. Leith\unskip,\iSLAC}
\mbox{V. Lia\unskip,\iMIT}
\mbox{C.Lin\unskip,\iMASS}
\mbox{M.X. Liu\unskip,\iYALE}
\mbox{X. Liu\unskip,\iUCSC}
\mbox{M. Loreti\unskip,\iPADO}
\mbox{A. Lu\unskip,\iUCSB}
\mbox{H.L. Lynch\unskip,\iSLAC}
\mbox{J. Ma\unskip,\iWASH}
\mbox{G. Mancinelli\unskip,\iRUTG}
\mbox{S. Manly\unskip,\iYALE}
\mbox{G. Mantovani\unskip,\iPERU}
\mbox{T.W. Markiewicz\unskip,\iSLAC}
\mbox{T. Maruyama\unskip,\iSLAC}
\mbox{H. Masuda\unskip,\iSLAC}
\mbox{E. Mazzucato\unskip,\iFERR}
\mbox{A.K. McKemey\unskip,\iBRUN}
\mbox{B.T. Meadows\unskip,\iCINC}
\mbox{G. Menegatti\unskip,\iFERR}
\mbox{R. Messner\unskip,\iSLAC}
\mbox{P.M. Mockett\unskip,\iWASH}
\mbox{K.C. Moffeit\unskip,\iSLAC}
\mbox{T.B. Moore\unskip,\iYALE}
\mbox{M.Morii\unskip,\iSLAC}
\mbox{D. Muller\unskip,\iSLAC}
\mbox{V.Murzin\unskip,\iMOSCOW}
\mbox{T. Nagamine\unskip,\iTOHO}
\mbox{S. Narita\unskip,\iTOHO}
\mbox{U. Nauenberg\unskip,\iCOLO}
\mbox{H. Neal\unskip,\iSLAC}
\mbox{M. Nussbaum\unskip,\iCINC}
\mbox{N.Oishi\unskip,\iNAGO}
\mbox{D. Onoprienko\unskip,\iTENN}
\mbox{L.S. Osborne\unskip,\iMIT}
\mbox{R.S. Panvini\unskip,\iVAND}
\mbox{C. H. Park\unskip,\iSOONG}
\mbox{T.J. Pavel\unskip,\iSLAC}
\mbox{I. Peruzzi\unskip,\iFRAS}
\mbox{M. Piccolo\unskip,\iFRAS}
\mbox{L. Piemontese\unskip,\iFERR}
\mbox{K.T. Pitts\unskip,\iOREG}
\mbox{R.J. Plano\unskip,\iRUTG}
\mbox{R. Prepost\unskip,\iWISC}
\mbox{C.Y. Prescott\unskip,\iSLAC}
\mbox{G.D. Punkar\unskip,\iSLAC}
\mbox{J. Quigley\unskip,\iMIT}
\mbox{B.N. Ratcliff\unskip,\iSLAC}
\mbox{T.W. Reeves\unskip,\iVAND}
\mbox{J. Reidy\unskip,\iMISSI}
\mbox{P.L. Reinertsen\unskip,\iUCSC}
\mbox{P.E. Rensing\unskip,\iSLAC}
\mbox{L.S. Rochester\unskip,\iSLAC}
\mbox{P.C. Rowson\unskip,\iCOLU}
\mbox{J.J. Russell\unskip,\iSLAC}
\mbox{O.H. Saxton\unskip,\iSLAC}
\mbox{T. Schalk\unskip,\iUCSC}
\mbox{R.H. Schindler\unskip,\iSLAC}
\mbox{B.A. Schumm\unskip,\iUCSC}
\mbox{J. Schwiening\unskip,\iSLAC}
\mbox{S. Sen\unskip,\iYALE}
\mbox{V.V. Serbo\unskip,\iSLAC}
\mbox{M.H. Shaevitz\unskip,\iCOLU}
\mbox{J.T. Shank\unskip,\iBU}
\mbox{G. Shapiro\unskip,\iLBL}
\mbox{D.J. Sherden\unskip,\iSLAC}
\mbox{K. D. Shmakov\unskip,\iTENN}
\mbox{C. Simopoulos\unskip,\iSLAC}
\mbox{N.B. Sinev\unskip,\iOREG}
\mbox{S.R. Smith\unskip,\iSLAC}
\mbox{M. B. Smy\unskip,\iCSU}
\mbox{J.A. Snyder\unskip,\iYALE}
\mbox{H. Staengle\unskip,\iCSU}
\mbox{A. Stahl\unskip,\iSLAC}
\mbox{P. Stamer\unskip,\iRUTG}
\mbox{H. Steiner\unskip,\iLBL}
\mbox{R. Steiner\unskip,\iADEL}
\mbox{M.G. Strauss\unskip,\iMASS}
\mbox{D. Su\unskip,\iSLAC}
\mbox{F. Suekane\unskip,\iTOHO}
\mbox{A. Sugiyama\unskip,\iNAGO}
\mbox{S. Suzuki\unskip,\iNAGO}
\mbox{M. Swartz\unskip,\iJHU}
\mbox{A. Szumilo\unskip,\iWASH}
\mbox{T. Takahashi\unskip,\iSLAC}
\mbox{F.E. Taylor\unskip,\iMIT}
\mbox{J. Thom\unskip,\iSLAC}
\mbox{E. Torrence\unskip,\iMIT}
\mbox{N. K. Toumbas\unskip,\iSLAC}
\mbox{T. Usher\unskip,\iSLAC}
\mbox{C. Vannini\unskip,\iPISA}
\mbox{J. Va'vra\unskip,\iSLAC}
\mbox{E. Vella\unskip,\iSLAC}
\mbox{J.P. Venuti\unskip,\iVAND}
\mbox{R. Verdier\unskip,\iMIT}
\mbox{P.G. Verdini\unskip,\iPISA}
\mbox{D. L. Wagner\unskip,\iCOLO}
\mbox{S.R. Wagner\unskip,\iSLAC}
\mbox{A.P. Waite\unskip,\iSLAC}
\mbox{S. Walston\unskip,\iOREG}
\mbox{J.Wang\unskip,\iSLAC}
\mbox{S.J. Watts\unskip,\iBRUN}
\mbox{A.W. Weidemann\unskip,\iTENN}
\mbox{E. R. Weiss\unskip,\iWASH}
\mbox{J.S. Whitaker\unskip,\iBU}
\mbox{S.L. White\unskip,\iTENN}
\mbox{F.J. Wickens\unskip,\iRAL}
\mbox{B. Williams\unskip,\iCOLO}
\mbox{D.C. Williams\unskip,\iMIT}
\mbox{S.H. Williams\unskip,\iSLAC}
\mbox{S. Willocq\unskip,\iMASS}
\mbox{R.J. Wilson\unskip,\iCSU}
\mbox{W.J. Wisniewski\unskip,\iSLAC}
\mbox{J. L. Wittlin\unskip,\iMASS}
\mbox{M. Woods\unskip,\iSLAC}
\mbox{G.B. Word\unskip,\iVAND}
\mbox{T.R. Wright\unskip,\iWISC}
\mbox{J. Wyss\unskip,\iPADO}
\mbox{R.K. Yamamoto\unskip,\iMIT}
\mbox{J.M. Yamartino\unskip,\iMIT}
\mbox{X. Yang\unskip,\iOREG}
\mbox{J. Yashima\unskip,\iTOHO}
\mbox{S.J. Yellin\unskip,\iUCSB}
\mbox{C.C. Young\unskip,\iSLAC}
\mbox{H. Yuta\unskip,\iAOMORI}
\mbox{G. Zapalac\unskip,\iWISC}
\mbox{R.W. Zdarko\unskip,\iSLAC}
\mbox{J. Zhou\unskip.\iOREG}

\it
  \vskip \baselineskip                   
  \vskip \baselineskip        
  \baselineskip=.75\baselineskip   
\iADEL
  Adelphi University, Garden City, New York 11530, \break
\iAOMORI
  Aomori University, Aomori , 030 Japan, \break
\iBOLO
  INFN Sezione di Bologna, I-40126, Bologna Italy, \break
\iBRI
  University of Bristol, Bristol, U.K., \break
\iBRUN
  Brunel University, Uxbridge, Middlesex, UB8 3PH United Kingdom, \break
\iBU
  Boston University, Boston, Massachusetts 02215, \break
\iCINC
  University of Cincinnati, Cincinnati, Ohio 45221, \break
\iCOLO
  University of Colorado, Boulder, Colorado 80309, \break
\iCOLU
  Columbia University, New York, New York 10533, \break
\iCSU
  Colorado State University, Ft. Collins, Colorado 80523, \break
\iFERR
  INFN Sezione di Ferrara and Universita di Ferrara, I-44100 Ferrara, Italy, \break
\iFRAS
  INFN Lab. Nazionali di Frascati, I-00044 Frascati, Italy, \break
\iILLI
  University of Illinois, Urbana, Illinois 61801, \break
\iJHU
  Johns Hopkins University, Baltimore, MD 21218-2686, \break
\iLBL
  Lawrence Berkeley Laboratory, University of California, Berkeley, California 94720, \break
\iLTU
  Louisiana Technical University - Ruston,LA 71272, \break
\iMASS
  University of Massachusetts, Amherst, Massachusetts 01003, \break
\iMISSI
  University of Mississippi, University, Mississippi 38677, \break
\iMIT
  Massachusetts Institute of Technology, Cambridge, Massachusetts 02139, \break
\iMOSCOW
  Institute of Nuclear Physics, Moscow State University, 119899, Moscow Russia, \break
\iNAGO
  Nagoya University, Chikusa-ku, Nagoya 464 Japan, \break
\iOREG
  University of Oregon, Eugene, Oregon 97403, \break
\iOXF
  Oxford University, Oxford, OX1 3RH, United Kingdom, \break
\iPADO
  INFN Sezione di Padova and Universita di Padova I-35100, Padova, Italy, \break
\iPERU
  INFN Sezione di Perugia and Universita di Perugia, I-06100 Perugia, Italy, \break
\iPISA
  INFN Sezione di Pisa and Universita di Pisa, I-56010 Pisa, Italy, \break
\iRAL
  Rutherford Appleton Laboratory, Chilton, Didcot, Oxon OX11 0QX United Kingdom, \break
\iRUTG
  Rutgers University, Piscataway, New Jersey 08855, \break
\iSLAC
  Stanford Linear Accelerator Center, Stanford University, Stanford, California 94309, \break
\iSOGA
  Sogang University, Seoul, Korea, \break
\iSOONG
  Soongsil University, Seoul, Korea 156-743, \break
\iTENN
  University of Tennessee, Knoxville, Tennessee 37996, \break
\iTOHO
  Tohoku University, Sendai 980, Japan, \break
\iUCSB
  University of California at Santa Barbara, Santa Barbara, California 93106, \break
\iUCSC
  University of California at Santa Cruz, Santa Cruz, California 95064, \break
\iUVIC
  University of Victoria, Victoria, B.C., Canada, V8W 3P6, \break
\iVAND
  Vanderbilt University, Nashville,Tennessee 37235, \break
\iWASH
  University of Washington, Seattle, Washington 98105, \break
\iWISC
  University of Wisconsin, Madison,Wisconsin 53706, \break
\iYALE
  Yale University, New Haven, Connecticut 06511. \break

\rm
%

\end{center}


\vfill
\eject

\clearpage  
 
\noindent
\begin{table} [ht]
\begin{center}
\begin{tabular}{|c|c|c|} \hline 
Jet label  & \# Tagged gluon jets & Purity\\
\hline
3 & 1140 & 94.4 $\%$ \\
2 & 155  & 90.1 $\%$ \\
1 & 34   & 73.1 $\%$ \\
\hline 
\end{tabular}
\end{center}
\caption{Estimated purities of the tagged gluon-jet samples.}
\label{dtagpure}
\end{table}

\begin{table} [ht]
\begin{center}
\begin{tabular}{|l|c|} \hline
QCD Calculation & $\chi^2$: $x_g$ (10 bins) \\
\hline
LO $m_b(m_Z)$ = 3 GeV/$c^2$ & 170 \\
NLO $m_b(m_Z)$ = 3 GeV/$c^2$ & 51 \\
PS $M_b$ = 5 GeV/$c^2$ & 18 \\
\hline
\end{tabular}
\end{center}
\caption{$\chi^2$ for the comparison of the QCD predictions with the corrected 
data.}
\label{chisq}
\end{table}

\clearpage

\begin{figure}[hbtp]
\begin{center}
\leavevmode
\epsfysize=15 cm.
\epsffile{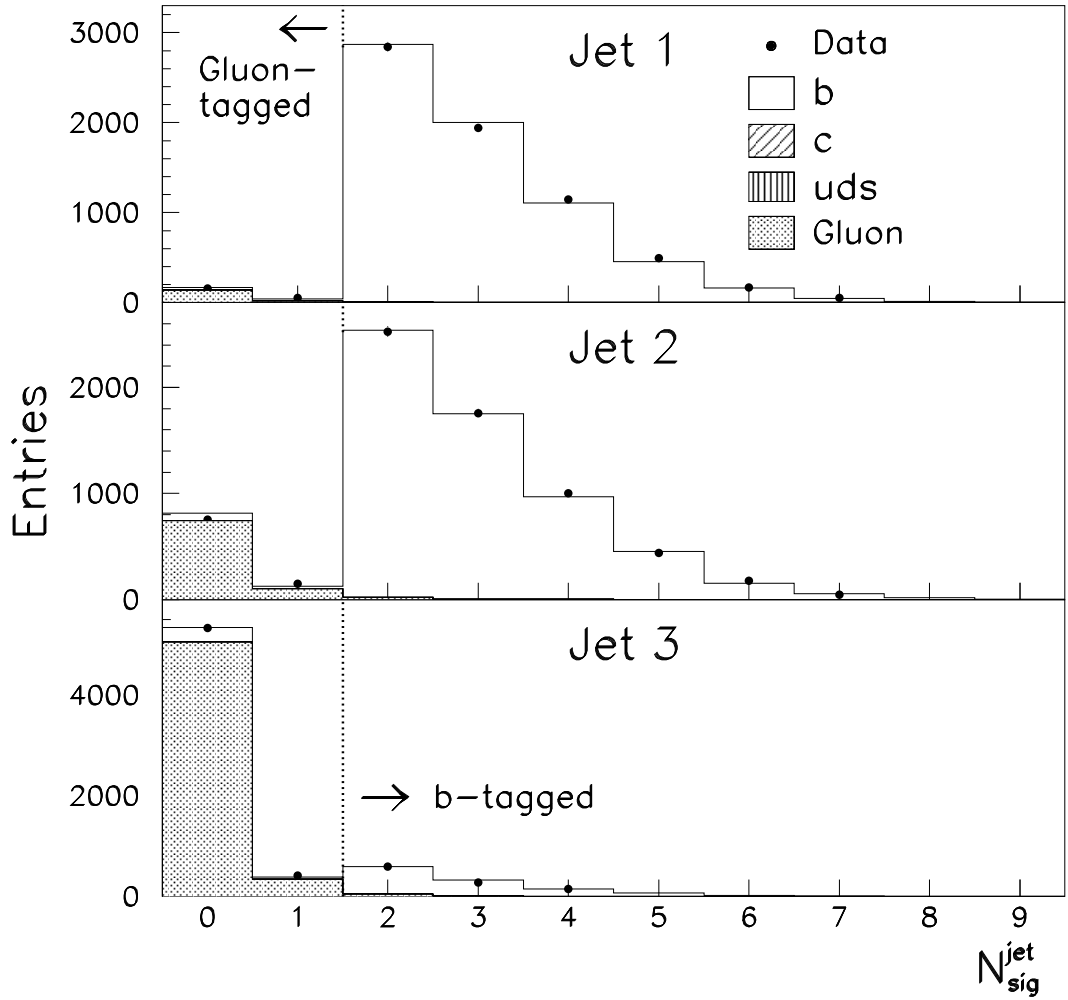}
\end{center}
\caption{The $N_{sig}^{jet}$ distributions for jets in \bbg-tagged events, 
labelled according to
jet energy (dots); errors are statistical. Histograms: 
simulated distributions showing jet flavour contributions.
}
\end{figure}

\begin{figure}[hbtp]
\begin{center}
\leavevmode
\epsfysize=15 cm.
\epsffile{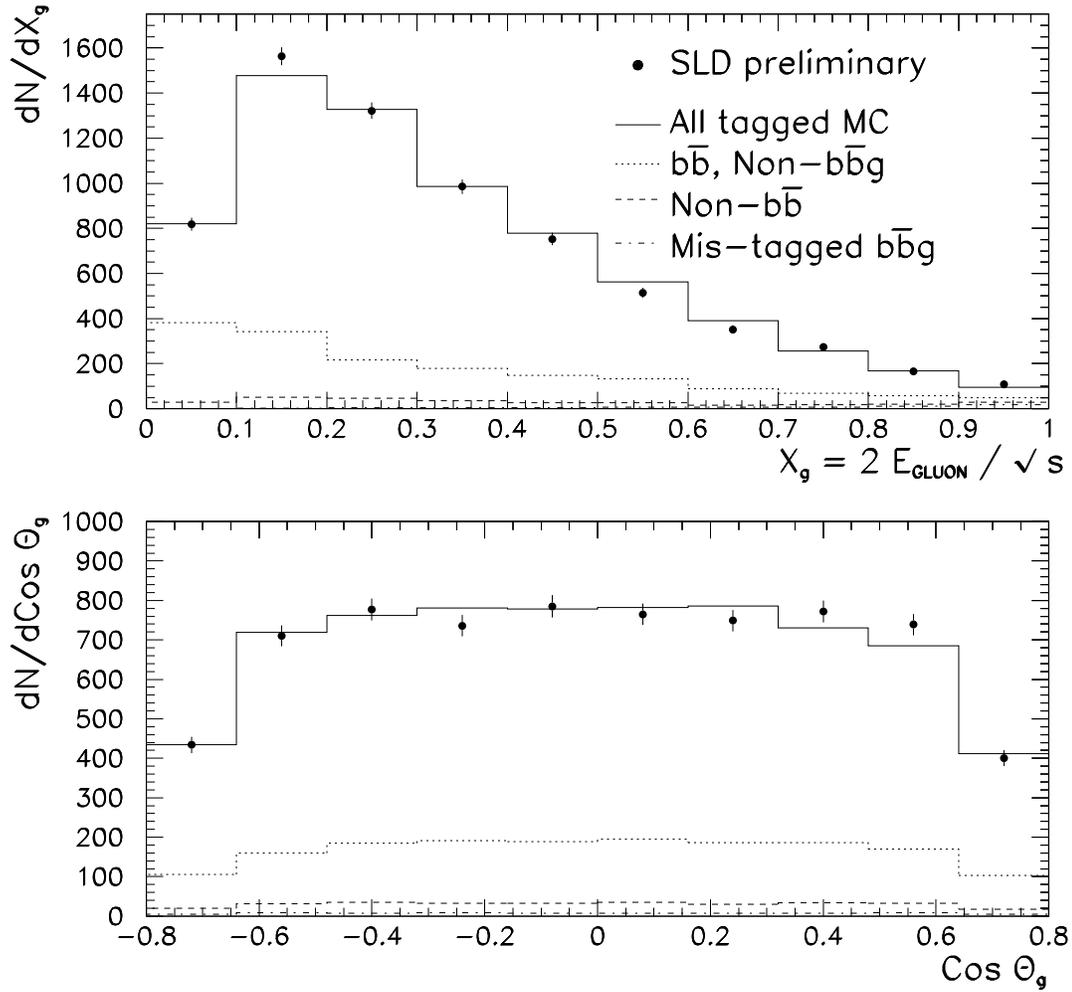}
\end{center}
\caption{Raw measured distributions of (a) $x_g$ and (b) cos$\theta_g$ (dots); 
errors are statistical. Histograms:
simulated distributions including background contributions.
}
\end{figure}

\begin{figure}[hbtp]
\begin{center}
\leavevmode
\epsfysize=15 cm.
\epsffile{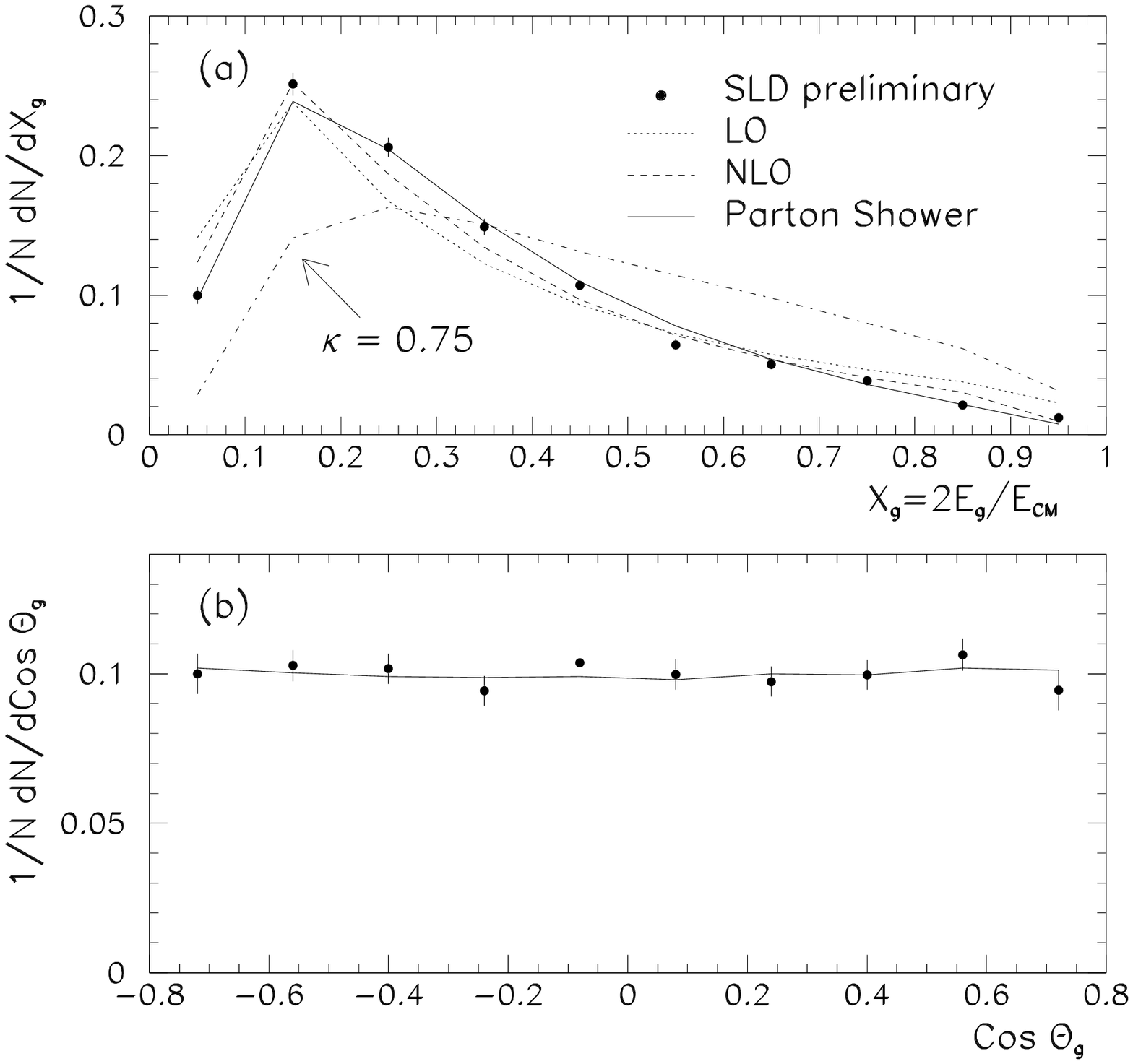}
\end{center}
\caption{Corrected distributions of (a) $x_g$ and (b) cos$\theta_g$ (dots);
errors are statistical.
Perturbative QCD predictions (see text) are shown as lines joining entries
plotted at the respective bin centers.
}
\end{figure}


\begin{thebibliography}{99}

\bibitem{threejets} 
See \eg S.L. Wu, Phys. Rept. {\bf 107} (1984) 59.\\
J. Ellis, M. K. Gaillard, and G. G. Ross, Nucl. Phys. {\bf B111} (1976) 253;
erratum: {\it ibid.} {\bf B130} (1977) 516.

\bibitem{BO} 
See \eg P.N. Burrows, P. Osland, \plb {\bf B400} (1997) 385.

\bibitem{quark} 
We do not distinguish between particle and antiparticle.

\bibitem{VXD2} C. J. S. Damerell {\it et al}., Nucl. Inst. Meth. {\bf A288} (1990) 236.

\bibitem{VXD3} K. Abe {\it et al}., Nucl. Inst. Meth. {\bf A400} (1997) 287.

\bibitem{electrow} See \eg, G. C. Ross, Electroweak Interactions and Unified Theories, 
Proc. xXXXI Rencontre
de Moriond, 16-23 March 1996, Les Arcs, Savoie, France, Editions Frontieres 
(1996), ed. J. Tran Thanh Van, p 481.

\bibitem{sldflav} 
SLD Collab., K. Abe \etal, \prd {\bf D59} (1999) 012002.

\bibitem{sldsymm} 
SLD Collab., K. Abe \etal, SLAC-PUB-8157 (1999); contributed to this conference.

\bibitem{sldbbg} 
SLD Collab., K. Abe \etal, SLAC-PUB-7920 (1999); subm. to \prd D.

\bibitem{arnd} 
W. Bernreuther, A. Brandenburg, P. Uwer, \prl {\bf 79} (1997) 189. \\
A. Brandenburg, P. Uwer, Nucl. Phys. {\bf B515} (1998) 279.

\bibitem{tom1} T. Rizzo,  Phys. Rev. {\bf D50} (1994) 4478, 
and private communications.

\bibitem{SLD} SLD Design Report, SLAC Report 273 (1984).

\bibitem{dervan} 
P.J. Dervan, Brunel Univ. Ph.D. thesis; SLAC-Report-523 (1998).

\bibitem{jade} 
JADE Collab., W. Bartel {\it et. al.}, Z. Phys. {\bf C33} (1986) 23.

\bibitem{jetset} T. Sj\"{o}strand, Comp. Phys. Commun. {\bf 82} (1994) 74.

\bibitem{tuning} P. N. Burrows, Z. Phys. {\bf C41} (1988) 375.\\
OPAL Collab., M. Z. Akrawy {\it el al}., {\it ibid}. {\bf C47} (1990) 505.

\bibitem{bdecay} 
SLD Collab., K. Abe {\it et al}., Phys. Rev. Lett. {\bf 79} (1997) 590.
 
\bibitem{bbbb} 
We expect less than 0.4\% of the selected sample to comprise events
of the type \ep\ra\qqg, with $g$\ra\bb. In the evaluation of the purity only 
true \bb\bb events were considered as signal \bbg events; 
\qq\bb events ($q\neq b$) were considered as backgrounds.

\bibitem{bmass} 
DELPHI Collab., P.~Abreu \etal, Phys.  Lett. {\bf B418} (1998) 430.\\
SLD Collab., K. Abe \etal, \prd {\bf D59} (1999) 012002.\\
A. Brandenburg \etal, SLAC-PUB-7915 (1999); subm. to Phys. Lett. B.\\
OPAL Collab., CERN-EP/99-045 (1999); subm. to Eur. Phys. J. C.

\bibitem{hwang} 
SLD Collab., K. Abe {\it et al}., Phys. Rev. {\bf D55}, (1997) 2533.

\vspace{-0.2cm}
\end{thebibliography}
\end{document}